\documentclass[]{JHEP3}

\usepackage{graphicx}
\usepackage{graphics}
\usepackage{epsfig}
\epsfclipon

\usepackage{epsfig,bm}
\usepackage{latexsym}
\usepackage{amssymb,amsmath}

\newcommand{\nco}{\newcommand}
\nco{\beq}{\begin{equation}} \nco{\eeq}{\end{equation}}
\nco{\beqa}{\begin{eqnarray}} \nco{\eeqa}{\end{eqnarray}}

\def\be{\begin{equation}}
\def\ee{\end{equation}}

\def\baray{\begin{eqnarray}}
\def\earay{\end{eqnarray}}

\nco{\sss}{\scriptscriptstyle} \nco{\dphi}{\varphi}
\nco{\lsim}{\mbox{\raisebox{-.6ex}{~$\stackrel{<}{\sim}$~}}}
\nco{\gsim}{\mbox{\raisebox{-.6ex}{~$\stackrel{>}{\sim}$~}}}

\def\IK{\relax{\rm I\kern-.20em K}}
\def\IM{\relax{\rm I\kern-.20em M}}

\def\lsim{\mbox{\raisebox{-.6ex}{~$\stackrel{<}{\sim}$~}}}
\def\gsim{\mbox{\raisebox{-.6ex}{~$\stackrel{>}{\sim}$~}}}
\def\sss{\scriptscriptstyle}

\title{Dynamics and Stability of Light-Like Tachyon Condensation}

\author{Neil Barnaby \\ Canadian Institute for Theoretical Astrophysics,
University of Toronto, 60 St.\ George St.\, Toronto, Ontario M5S 3H8 Canada \\
Email: \email{barnaby@cita.utoronto.ca}}

\author{David J. Mulryne \\ Centre for Theoretical Cosmology,\\Department of Applied Mathematics and Theoretical Physics, Wilberforce Road, Cambridge, CB3 0WA, UK \\ Email: \email{D.Mulryne@damtp.cam.ac.uk}}

\author{Nelson J. Nunes \\ Institute f\"{u}r Theoretische Physik, Philosophenweg 16, D-69120 Heidelberg, Germany \\
Email: \email{n.nunes@thphys.uni-heidelberg.de}}

\author{Patrick Robinson \\ Canadian Institute for Theoretical Astrophysics,
University of Toronto, 60 St.\ George St.\, Toronto, Ontario M5S 3H8 Canada }

\preprint{HD-THEOP-08-25, DAMTP-2008-102} 
\abstract{ 

Recently, Hellerman and Schnabl considered the dynamics of unstable D-branes in the background of a linear dilaton.  Remarkably, they were able to construct light-like tachyon solutions which interpolate
smoothly between the perturbative and nonperturbative vacua, without undergoing the wild oscillations that plague time-like solutions.  In their analysis, however, the full structure of the initial value problem for the nonlocal dynamical equations 
was not considered.  In this paper, therefore, we reexamine the nonlinear dynamics of 
light-like tachyon condensation using a combination of numerical and 
analytical techniques.  We find that for the $p$-adic string  the 
monotonic behaviour obtained previously relied on a special choice of initial conditions 
near the unstable maximum.  
For generic initial conditions the wild oscillations come back to haunt us.  
Interestingly, we find an ``island of stability'' in initial condition space 
that leads to sensible evolution at late times.  For the string field theory case, on the other hand, we find that the evolution is \emph{completely stable} for generic choices of initial data.  
This provides an explicit example of a string theoretic system that admits infinitely many initial data but is nevertheless nonperturbatively stable.  Qualitatively similar dynamics are obtained 
in nonlocal cosmologies where the Hubble damping plays a role very analogous to the dilaton gradient.

}

\keywords{differential equations of infinite order, string field theory, $p$-adic strings}

\begin{document}

\section{Introduction}

Nonlinear theories with infinitely many derivatives have come to play an increasingly important role in theoretical physics;  attracting interest both from string theorists and cosmologists.  The string theory application
of such equations that has stimulated the most interest is that of understanding the dynamics of tachyon condensation in string field theory (SFT) \cite{sft,sft2}
and also in related toy models, such as the $p$-adic string \cite{padic_st}.  The dynamical process of  the tachyon field rolling from the unstable maximum of its potential to
the true vacuum is expected to give a time-dependent description of the decay of unstable D-brane configurations in string theory (see \cite{sen_rev} and references therein).  A long-standing puzzle has 
been the observation that generic tachyon solutions in flat space do not roll monotonically from the unstable maximum of the potential to the true vacuum of the theory, 
as would be expected on physical grounds.  Rather, the tachyon field undergoes wild oscillations at late times \cite{zwiebach}; a manifestation of the Ostrogradski instability \cite{ostrogradski}
that plagues higher derivative theories (see \cite{1/r} and \cite{woodard} for a more modern discussion). The string theoretic interpretation of this peculiar behaviour is a 
subtle problem and there are a number of proposed explanations in the literature (discussed in more detail below).  (See \cite{sft3} for further discussion of tachyon dynamics in SFT
and related theories and see \cite{sft4,sft5} for significant progress in understanding the vacuum structure of SFT.)

The wild oscillations of the tachyon field arise because the higher derivative structure of SFT allows for extra (more than two) initial data in the solutions of the field equations.
These extra data can be interpreted as a tower of new physical states, in addition to the usual tachyonic excitation, which contribute to the kinetic energy with indefinite sign \cite{pais}.  In 
the case of SFT and the $p$-adic string, these spurious extra states have complex mass-squared and behave as an admixture of ghost and non-ghost field (we will refer to such excitations
as ghost-like or ``quintom'' in the text).  The presence of extra ghost (or ghost-like) degrees of freedom is a rather generic problem for higher derivative theories; see \cite{woodard} 
for a detailed discussion.  In any theory with negative kinetic energy the Hamiltonian is unbounded from below and the system will become arbitrarily excited at late times for generic choices
of initial data.  The kind of peculiar time dependence obtained for rolling tachyon solutions in flat space is quite typical of higher derivative instabilities.

Recently, Hellerman and Schnabl \cite{hs} made significant progress in studying the dynamics of brane decay by explicitly constructing tachyon solutions that roll smoothly from the perturbative vacuum to the true vacuum, as 
a function of light-cone time, $x^{+}$.  This progress relied on turning on a linear dilaton background that violates energy conservation and provides a source of friction for the tachyon dynamics.  Hellerman 
and Schnabl considered light-like tachyon profiles in SFT, $p$-adic string theory and also vacuum string field theory (VSFT), finding similar dynamics in all three cases.  Despite their significant result, however, the analysis 
of  \cite{hs} does not explicitly consider the role of initial conditions near the unstable maximum.  Without a complete understanding of the initial value problem it is impossible to assess the stability of such solutions.
Here we point out that the solutions presented in  \cite{hs} result from the choice of one particular kind of initial condition 
from an infinite set of possibilities.  Furthermore, we will show that for the cases of the $p$-adic string and VSFT the monotonic behaviour obtained by Hellerman and Schnabl is not generic in the sense that it relies on the special 
choice of initial conditions made.  For more general choices of initial conditions the wild oscillations associated with the Ostrogradski instability come back to haunt us.  On the other hand, we will show that the equation obtained 
in \cite{hs} for SFT in a level zero truncation leads to completely stable dynamics, {\it even when the extra initial data are taken into account!}  This behaviour is confirmed by fully nonlinear numerical simulations, and provides a 
remarkable example of an interacting nonlocal theory that admits infinitely many initial data but whose evolution is nevertheless completely stable.  These surprising dynamics arise because the friction coming from the dilaton gradient 
efficiently damps out the oscillations of the tachyon field.  

A final surprise awaits us.  Further studying the $p$-adic and VSFT case, we discover that despite the instability associated with generic initial 
conditions, the set of initial conditions leading to stable evolution is not of measure zero.  This implies that even in these cases there is an ``island of stability'' in the initial condition space which leads to a well behaved evolution.

Nonlocal theories motivated by SFT have recently attracted interest also from cosmologists \cite{phantom1}-\cite{mulryne} due to a wide array of novel cosmological behaviours.
Of particular interest are recent efforts to construct inflationary solutions in nonlocal theories \cite{pi}-\cite{mulryne}.  The first realisation of nonlocal inflation was in the context of $p$-adic inflation \cite{pi}.  This model
can support inflation even when the potential is naively very steep, a behaviour that was found to be a rather generic feature of nonlocal
inflation in \cite{lidsey} (see also \cite{ng1}) and was verified using numerical analysis in \cite{mulryne}.  Moreover, nonlocal inflation is also one of the rare models that predicts large non-gaussianity in the cosmic microwave 
background \cite{ng1, ng2}.  Here we point out that the nonlinear dynamics of nonlocal inflation are strikingly similar to the light-like tachyon models considered by Hellerman and Schnabl.  Hence we expect that our 
analysis should have implications also for the stability of nonlocal cosmologies.

During the course of our investigation we uncover a number of results of general interest to the study of nonlinear theories with infinitely many derivatives.
For instance, we will argue that the recipe of mixing friction and constraints on the initial conditions provides a very generic prescription for constructing stable solutions in infinite
order theories.  We believe that this work should be useful in guiding future searches for stable solutions in SFT (and similar theories).  Moreover, our analysis clarifies a number of issues concerning the mathematical structure
of nonlinear infinite order differential equations.  In \cite{niky} the initial value problem for linear constant coefficient equations with infinitely many derivatives was studied and a formalism was developed to exhaustively
count initial data.\footnote{See also \cite{calcagni}-\cite{IVP} different approaches to the initial value problem and \cite{math} for mathematical analysis of $p$-adic and string field equations.}  These results were generalised
to the case of variable coefficient equations (such as those that arise when studying nonlocal cosmological perturbation theory) in \cite{niky2}.  However, the nonlinear problem presents a number of questions which
could not be addressed in \cite{niky,niky2}.  This work represents progress towards a complete understanding of nonlinear equations with infinitely many derivatives.

The organisation of this paper is as follows.  In section
\ref{light_padic} we study the nonlinear dynamics of the light-like 
$p$-adic string tachyon in a linear dilaton background, presenting 
non-linear analytic solutions which fully take into account the
freedom in fixing initial conditions.  In section \ref{light_SFT}
we study the analogous dynamics at level zero truncation in SFT.  In
this case analytical solutions are not available so we must turn to a numerical analysis.
In section \ref{numerics} we describe our formalism for solving infinite order differential
equations numerically, before applying our numerical approach to study the light-like 
SFT tachyon with generic initial conditions in section \ref{SFT_num_sec}. In section \ref{cosmo_sec} we 
comment on the stability
of nonlocal cosmologies. We briefly review some proposed stringy interpretations
of the wild oscillations in rolling tachyon solutions in section \ref{wild_sec}.  Finally, in section \ref{concl_sec} we conclude.

\section{Light-Like Tachyon Condensation in $p$-adic String Theory}
\label{light_padic}

\subsection{Set-Up and Equation of Motion}

We begin our investigation of the dynamics of light-like tachyon condensation by studying $p$-adic string theory \cite{padic_st} coupled to a linear
dilaton profile.  We employ the action proposed in \cite{hs}
\begin{equation}
\label{Spadic}
  S = \frac{1}{g_p^2}\int d^Dx\, e^{-\Phi}\,\left[-\frac{1}{2} (p^{-\alpha'\Box/2} \phi)^2 + \frac{1}{p+1}\phi^{p+1}\right]\,,
\end{equation}
where $p$ is a prime number that characterises the world-sheet coordinates, $\alpha' = m_s^{-2}$ (with $m_s$ the string mass), $g_p$ is related to the open string coupling and 
$\phi$ is the (dimensionless) tachyon field.
Although (\ref{Spadic}) is not meant to be construed as a realistic model of string theory, we will spend some time studying this theory because it is analytically tractable and
provides an excellent playground for studying the nonlinear dynamics of infinite order theories.  Following \cite{hs} we will work in terms of light-cone coordinates 
$x^{\pm} = (x^0 \pm x^1)/\sqrt{2}$ so that the metric takes the form $ds^2 = -(dx^0)^2 + (dx^1)^2 + d\vec{y}\cdot d\vec{y} = -2dx^{+} dx^{-} + d\vec{y}\cdot d\vec{y}$.  The $D-2$ 
transverse coordinates $\vec{y}$ will play no role in the ensuing analysis.  The dilaton background is taken as
\begin{equation}
\label{dilaton}
  \Phi(x) = V_\mu x^\mu = -V^{+} x^{-} - V^{-} x^{+} + \vec{V}\cdot \vec{y}\,.
\end{equation}
For a light-like field $\phi = \phi(x^{+})$ one obtains the equation of motion
\begin{equation}
\label{eom_padic}
  \phi(x^{+} + \alpha' V^{+} \ln p ) = p^{\alpha' V^2 /2 }\phi^p(x^{+})\,.
\end{equation}
Note that this finite difference equation can trivially be re-written as a pseudo-differential
equation
\begin{equation}
\label{eom_padic2}
  p^{\alpha' V^{+} \partial_{+}} \phi(x^{+}) = p^{\alpha' V^2 /2 }\phi^p(x^{+})\,.
\end{equation}
This equation admits constant solutions $\phi = p^{-\alpha' V^2 / \left[2(p-1)\right]}$ and $\phi = 0$.  The former corresponds to
the unstable maximum of the potential (physically the state with a space-filling brane) while the latter is the true minimum (physically the 
nonperturbative vacuum with no D-brane).

The action (\ref{Spadic}) was motivated by the striking similarity Eq.~(\ref{eom_padic}) bears to the equation of motion
one derives for the tachyon in VSFT (with light-like ansatz and linear dilaton background).  Equation (\ref{eom_padic2}) is also identical (up 
to a re-scaling of the field and space-time coordinates) to the friction dominated equation 
for the inflaton dynamics in $p$-adic inflation \cite{pi}.\footnote{Setting $V^2 = 0$ and replacing $x^{+} \rightarrow t$, $V^{+} \rightarrow 3 H_0 / 2$
one obtains exactly the equation that was studied in section 4.3 of \cite{pi}.}  Due to these similarities we expect our analysis of (\ref{eom_padic2}) to
have relevance to VSFT and $p$-adic inflation also.

\subsection{Exact Analytic Solution}

Remarkably, equation (\ref{eom_padic}) admits an exact nonperturbative solution:
\begin{equation}
  \phi(x^{+}) = p^{-\alpha' V^2 / \left[2(p-1)\right]} \exp\left[- e^{x^{+}/(\alpha' V^{+})} F(x^{+}) \right] \label{phi}\,, \\
\end{equation}
where $F(x^{+})$ is an arbitrary smooth periodic function satisfying
\begin{equation}
\label{F}
  F(x^{+}) = F\left(x^{+} + \alpha' V^{+} \ln p\right)\,.
\end{equation}
It follows that $F$ can be decomposed into a Fourier series as
\begin{equation}
\label{F_series}
  F(x^{+}) = a_0 + \sum_{n=1}^{\infty} a_n \cos\left(\frac{2\pi n x^{+}}{\alpha' V^{+}\ln p}\right)  + \sum_{n=1}^{\infty} b_n \sin\left(\frac{2\pi n x^{+}}{\alpha' V^{+}\ln p}\right) \,.
\end{equation}
{ It is clear that the solution contains an infinite number of free parameters $\{a_n, b_n\}$ which allow us to fix the state of the solution at some initial time $x^{+}_{\rm i}$ (we will
usually set $x^{+}_{\rm i} \equiv 0$ in subsequent analysis). Hence an infinite number of initial conditions must be specified to give us a unique solution. As we will see, the first term in (\ref{F_series}) 
(proportional to $a_0$) is associated with the tachyonic excitation while the the oscillatory terms   (proportional to $a_n,b_n$ with $n>0$) are associated with 
the presence of ghosts in the theory, since these modes contribute with indefinite sign to the total kinetic energy.  We will refer to the $n \not= 0$ states as ghost-like (or quintom) modes.

Let us discuss the dynamics of the solution (\ref{phi}) in more detail. For any choices of $a_n$ and $b_n$, 
the solution rolls from the unstable maximum in the asymptotic past $x^{+} \rightarrow -\infty$.  
For $a_0>0$ and $a_n = b_n=0$ for $n>0$ the field rolls monotonically from the 
unstable maximum to the true vacuum of the theory.  This solution corresponds to turning on only the  
tachyon in the initial state and in this case our solution (\ref{phi}) matches 
the one derived in \cite{hs} (also identical to the solution obtained in \cite{pi} in a somewhat different context).  

On the other hand, considering $a_0=0$ and taking any of the $a_n$ or $b_n$ to be non-zero leads to a wildly oscillating solution which never settles to the minimum, but rather the amplitude of the oscillations grow in an unbounded 
manner (similarly to the solutions of (\ref{Spadic}) in the background of a constant dilaton \cite{zwiebach}). This is not unexpected, since these modes of excitation are ghost-like and lead to vacuum instability.

It is extremely interesting and indeed unexpected, however, to see what occurs if we choose to turn on some 
admixture of ghost excitation together with the well behaved tachyon in the initial state. Let us again take $a_0 > 0$.  
In this case we can choose the origin of time such that $a_0=1$ without loss of generality.  Now let us arbitrarily take some of the $a_n,b_n$ to be 
non-zero, but sufficiently small in a sense which we will quantify shortly.  In this case the solution again rolls away from the unstable maximum, but as it rolls 
down the potential the field undergoes some small oscillations, though it still settles down 
to the minimum at late times.  
Quantitatively, the condition  
\begin{equation}
\label{cond1}
  F(x^{+}) > 0
\end{equation} 
is sufficient to ensure that $\phi(x^{+}) \rightarrow 0$ as $x^{+} \rightarrow \infty$.  Let us now consider increasing $a_n,b_n$ so that (\ref{cond1}) is violated.  In this case the ghost excitations become dominant and 
$\phi$ oscillates wildly at late 
times.  Note that these initial conditions are physically reasonable because $\phi$ is at the false 
vacuum in the asymptotic past $x^{+} \rightarrow -\infty$.  In Fig. \ref{phi_fig} we illustrate the different 
behaviours of the solution (\ref{phi}), which we have just discussed, for some representative choices of $a_n,b_n$.}

\EPSFIGURE{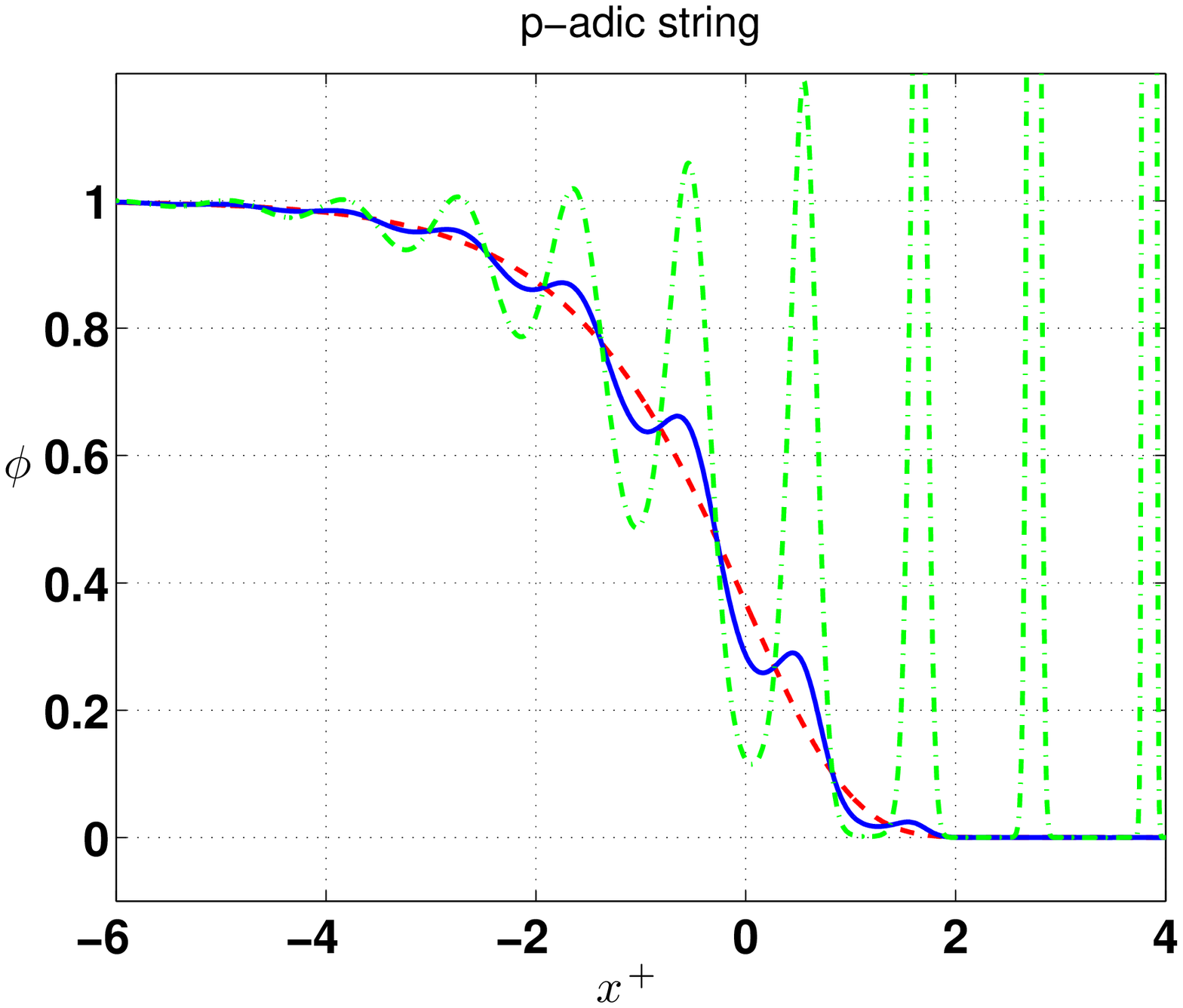,width=4in}{The solution $\phi(x^{+})$, Eq.~(\ref{phi}),  as a function of $x^{+}$.  The curves are $a_1 = 0$ (dashed line), $a_1=0.25$ (solid line), $a_1=1.1$ (dash-dotted line) with all other $a_n$
and $b_n$ vanishing and $a_0 = 1$.  For illustration we have set $p=3$ and $V^2 = 0$.  The $x^{+}$ axis is measured in units of $\alpha'V^{+}$.\label{phi_fig}}

Let us comment on the structure of the initial value problem for (\ref{eom_padic2}), which is closely related to the (in)stability of the theory.  In \cite{niky} a formalism was presented for exhaustively counting 
the initial data of linear, infinite order equations (see \cite{niky2} for a generalisation to variable coefficient equations).  One could use this formalism to (\ref{eom_padic2}) in a perturbative expansion around 
$\phi = p^{-\alpha' V^2 / \left[2(p-1)\right]}$.  The solution so obtained reproduces the terms in a small-$u$ expansion of (\ref{phi}).  Using the results of \cite{niky} then proves that the solution (\ref{phi})
provides a complete solution of the initial value problem near the false vacuum.  By continuity, we expect that this solution is complete also nonperturbatively.

\subsection{Behaviour of the Hamiltonian}

Using our exact solution (\ref{phi}) it is straightforward to explicitly construct the energy density as a function of $x^{+}$ (this is not simply a constant because the dilaton gradient violates time translation
invariance).  For simplicity we consider $V^{-} = \vec{V} = 0$ so that $\Phi = -V^{+} x^{-}$ and $V^2= 0$ although we do not expect that our qualitative results depend on this restriction in any crucial way.  
The nontrivial components of the stress tensor are \cite{hs}
\begin{eqnarray}
   T_{+-} &=& \frac{e^{-\Phi}}{g_p^2} \left[-\frac{1}{2}\left(p^{-\alpha'\Box/2}\phi\right)^2 + \frac{\phi^{p+1}}{p+1}\right]\,, \\
  T_{++} &=& -\frac{\alpha'\ln p}{g_p^2} e^{-\Phi} \int_0^1 d\zeta \partial_{+}\left[\phi(x^{+} + V^{+}\alpha'\ln p \zeta)\right]\partial_{+}\phi(x^{+})\,.
\end{eqnarray}
Evaluating these on the solution (\ref{phi}) we have
\begin{eqnarray}
  -g_p^2 e^{\Phi}T_{+-} &=& \left[ \frac{e^{-2 e^{x^{+}/(\alpha'V^{+})}F(x^{+}) }}{2} - \frac{e^{-(p+1) e^{x^{+}/(\alpha'V^{+})}F(x^{+}) }}{p+1} \right] \label{Tpm}\,, \\
  -g_p^2 e^{\Phi}  T_{++} &=& \frac{e^{2x^{+} / (\alpha'V^{+})}}{\alpha' (V^{+})^2} \exp\left[-e^{x^{+}/(\alpha'V^{+})}F(x^{+})\right]\left[F(x^{+}) + \alpha'V^{+}F'(x^{+})\right] \nonumber \\
 && \int_{1}^{p}dz e^{-z e^{x^{+}/(\alpha'V^{+})}F(x^{+} + \alpha'V^{+}\ln z)}\left[ F(x^{+} + \alpha'V^{+}\ln z) +\right. \nonumber \\
  &&\left.  \alpha'V^{+}  F'(x^{+} + \alpha'V^{+}\ln z) \right] \label{T++}\,,
\end{eqnarray}
where in evaluating $T_{++}$ we have switched variables $z = p^{\zeta}$ in the integration.  Remembering that $T_{\mu\nu}$ transforms as a tensor the energy density is
\begin{equation}
\label{H}
  \rho = -\frac{1}{2}T_{++} - T_{+-}
\end{equation}
Since (\ref{Tpm}) could be written as $T_{+-} = - e^{-\Phi} U(\phi)$ where $U(\phi) = -\left.\mathcal{L}\right|_{\Box=0}$ it is natural to associate the second term in (\ref{H}) with the potential
energy and the first term with the kinetic energy.  It can be readily verified that the kinetic term has indefinite sign and that the presence of some non-zero $a_n,b_n$ (with $n>1$) in (\ref{F_series})
leads to negative contributions to the kinetic energy.

\EPSFIGURE{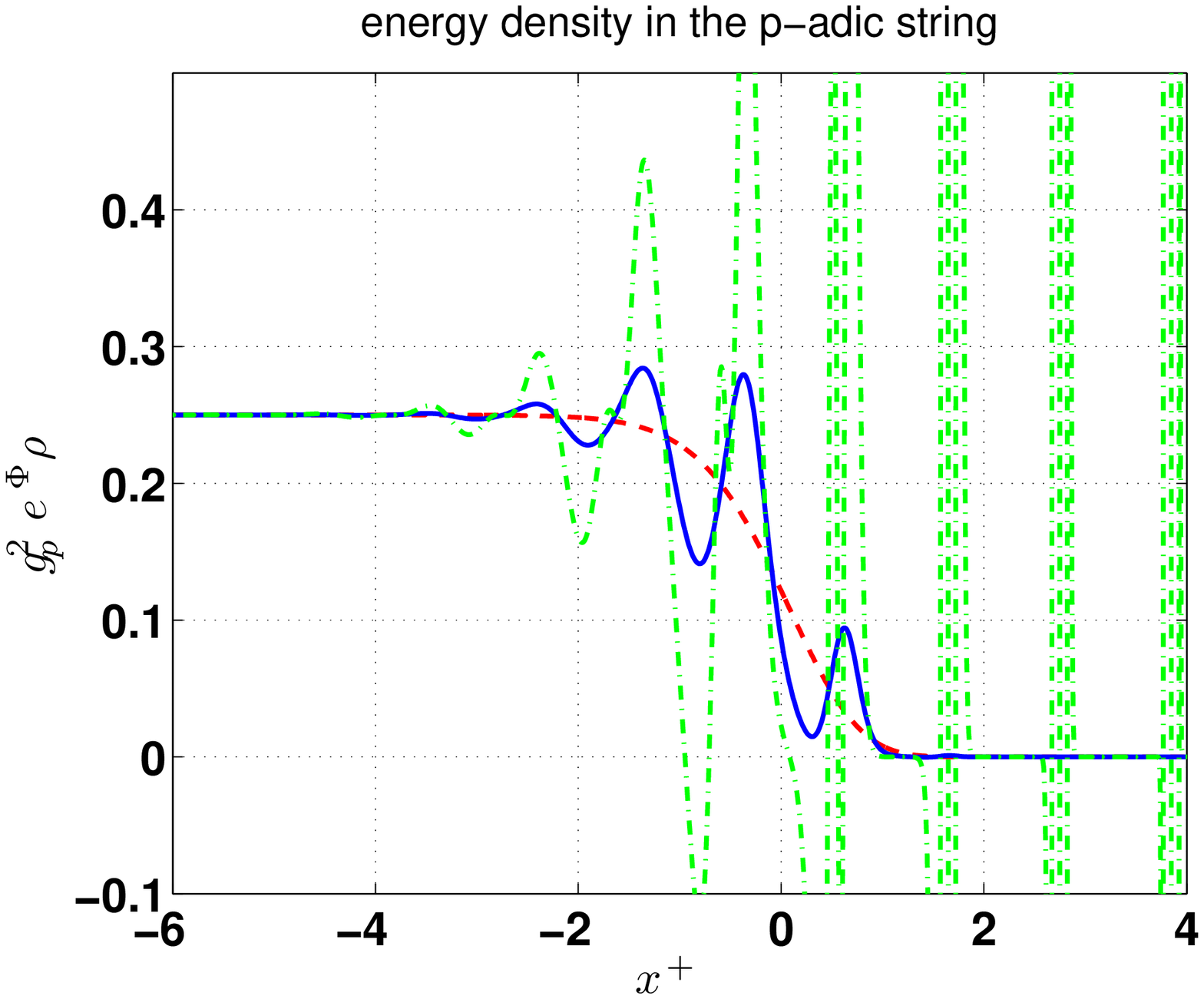,width=4in}{The energy density $g_p^2 e^{\Phi}\rho$ as a function of $x^{+}$ for the same values of $p$ and $a_n,b_n$ used in fig.\ \ref{phi_fig}.  The $x^{+}$ axis
is measured in units of $\alpha'V^{+}$ and, for illustration, we have set $V^{+}=1$ in these units.\label{H_fig}}

The behaviour of the energy density as a function of $x^{+}$ is plotted in Fig.~\ref{H_fig}.
In the case $a_n = b_n = 0$ for $n>0$ (the case considered by \cite{hs}) we have monotonic behaviour of $\rho(x^{+})$.  Turning on the ghost-like $n>0$ modes the oscillations of $\phi$ lead to oscillations in $\rho$.
If the contamination of ghost modes in the initial state is sufficiently large then these oscillations cause $\rho(x^{+})$ to cross zero and one would expect recollapse if gravity had been included.
From the previous discussion of $\phi(x^{+})$ one might expect that (\ref{cond1}) is a sufficient condition to keep the kinetic energy positive for all  $x^{+}$.  However, detailed examination of (\ref{H}) reveals 
that the actual condition is somewhat stronger:
\begin{equation}
\label{cond2}
  F(x^{+}) + \alpha' V^{+} F'(x^{+}) > 0\,.
\end{equation}
The fact that (\ref{cond2}) is a stronger constraint than (\ref{cond1}) means that we can construct solutions where $\rho(x^{+})$ crosses zero without prohibiting $\phi(x^{+})$ from settling down to the minimum at late times.

\subsection{The Island of Stability}

The instability of the theory (\ref{eom_padic2}) is not surprising.  This equation admits infinitely many initial conditions and hence Ostrogradski's theorem implies that the dynamics should be
generically unstable.  Of course it may be possible to construct well-behaved solutions by carefully choosing initial conditions, however, one would expect the set of initial conditions 
leading to non-pathological evolution to be a set of measure zero.  This intuition is based on the observation that the Ostrogradski Hamiltonian is linear in the unstable directions.
However, this intuition is incorrect in our case: the set of initial conditions leading to sensible evolution is \emph{not} measure zero.  Equation (\ref{eom_padic2}) allows for a certain amount of ghost-like contamination in
the initial conditions (quantified by taking $a_n,b_n$ to be non-zero) without prohibiting the tachyon from settling down to the vacuum at late times.  This observation implies that there is an ``island of stability'' in initial
condition space.\footnote{The expression ``island of stability'' is taken from ref.\ \cite{smilga} where similar behaviour was claimed for a finite order higher derivative theory.}  This (perhaps) surprising
behaviour arises because the dilaton friction can efficiently damp out the unstable growth of the ghost modes as long as they do not contaminate the initial state too strongly.

It is worth emphasising that the island of stability is \emph{not} a perturbative artifact.  Our solution (\ref{phi}) is exact at the fully nonlinear level and, as we have argued, it provides a full solution of the initial
value problem for the theory (\ref{eom_padic2}).  It is tempting to argue that a theory like (\ref{eom_padic2}) could be phenomenologically viable once the initial data are suitably restricted.  Such a restriction
on the data could be elegantly implemented using the contour deformation prescription advocated in \cite{niky}.\footnote{The solution (\ref{phi}) provides an explicit example showing that this prescription can be
applied at the fully nonlinear level.} We would, however, like to emphasise two caveats.  First, our analysis is purely classical.  Quantum mechanically a solution which starts in the island of stability might be
able to tunnel to an unstable configuration, perhaps though a negative tension instanton solution similar to those constructed in \cite{neg_tension}.  
Second, we should emphasise that the stability of the solution (\ref{phi}) is closely tied to the light-like ansatz $\phi = \phi(x^{+})$.  On scales small compared to the dilaton gradient one can treat $\Phi$
as a constant and (\ref{Spadic}) should have all the usual instabilities \cite{woodard} and wild oscillations \cite{zwiebach}.

It is interesting to compare our findings to the case of closed string tachyon condensation.  Purely light-like closed string tachyon solutions are stable against small perturbations of the initial conditions
(in the sense that small perturbations of the initial conditions lead to qualitatively similar behaviour at late times) \cite{closed1,closed2}.  On the other hand, \cite{closed3} discussed the  possibility of qualitative
change of the solution under finite changes of the initial conditions.

\section{Light-Like Tachyon Condensation in String Field Theory}
\label{light_SFT}

\subsection{Set-Up and Equation of Motion}

We now consider light-like tachyon condensation in the more realistic context of string field theory at level zero truncation.  The action derived in \cite{hs}
for the SFT tachyon in the linear dilaton background (given by Eq.~(\ref{dilaton})) is
\begin{equation}
\label{Ssft}
  S = -\frac{1}{g_o^2}\int d^D x\, e^{-\Phi}\,\left[ \frac{\alpha'}{2}(\partial \phi) - \frac{1}{2}\phi^2 + \frac{1}{3}K^{-3 + \alpha' V^2}\left(K^{-\alpha' \Box} \phi \right)^3 \right]\,,
\end{equation}
where $g_o$ is the open string coupling constant and the constant is $K = 4/3^{3/2}$.  Imposing again a light-like ansatz on the tachyon field $\phi = \phi(x^{+})$ we obtain the equation
of motion
\begin{equation}
\label{eom_SFT}
  \left(\alpha'V^{+}\partial_{+} - 1\right)\phi(x^{+}) + \frac{1}{K^3}\phi^2(x^{+} + 2\alpha' V^{+}\ln K) = 0\,.
\end{equation}
This can be re-written in pseudo-differential form as
\begin{equation}
\label{eom_SFT2}
  \left(\alpha'V^{+}\partial_{+} - 1\right)K^{-2\alpha'V^{+}\partial_{+}} \phi(x^{+}) + \frac{1}{K^3}\phi^2(x^{+}) = 0\,.
\end{equation}
This equation admits constant solutions $\phi = 0$ and $\phi = K^3$.  The former corresponds to the unstable maximum while the latter
is the true vacuum.

Note that, as in the $p$-adic case, equation (\ref{eom_SFT2}) is identical to what one would have for the time-like SFT tachyon in a de Sitter background in the limit of very large Hubble scale
where one could take $\Box = -\partial_t^2 - 3 H \partial_t \cong -3H\partial_t$.  Hence, in some (limited) sense the dilaton gradient acts like an infinite source of Hubble friction.

\subsection{Perturbative Analysis}
\label{pert_subsec}

Sadly, the equation (\ref{eom_SFT2}) does not seem to admit a simple exact analytic solution analogous to (\ref{phi}).  However, we can study the dynamics of this theory analytically using perturbation theory.
Near the false vacuum we can write
\begin{equation}
  \phi = 0+\delta\phi
\end{equation}
and linearize (\ref{eom_SFT2}) in $\delta \phi$ to obtain
\begin{equation}
   \left(\alpha'V^{+}\partial_{+} - 1\right)\delta\phi = 0\,.
\end{equation}
This equation admits a single growing mode
\begin{equation}
\label{single_grow}
  \delta\phi = a_0 e^{x^{+} / (\alpha'V^{+})}
\end{equation}
with $a_0$ constant.  This solutions describes the usual tachyonic instability near the false vacuum.

Near the true vacuum we take
\begin{equation}
  \phi = K^3 + \delta\phi
\end{equation}
and linearize (\ref{eom_SFT2}) in $\delta \phi$ to obtain
\begin{equation}
   \left(\alpha'V^{+}\partial_{+} - 1 + 2 K^{2\alpha'V^{+}\partial_{+}}\right)\delta\phi = 0\,.
\end{equation}
This equation belongs to the class studied in \cite{niky} where the generatrix is $f(s) = \alpha V^{+} s - 1 + 2K^{\alpha'V^{+} s}$.  The full solution is 
\begin{equation}
\label{vac_mode}
  \delta\phi = \sum_n a_n e^{s_n x^{+}}\,,
\end{equation}
where $a_n$ are arbitrary complex numbers and 
\begin{equation}
\label{vac_freq}
  \alpha' V^{+} s_n = 1 - \frac{1}{2\ln K} W_n\left[4 K^2 \ln K \right]\,.
\end{equation}
Here $W_n$ are the branches of the Lambert-W function and $n$ runs over all integer values (both positive and negative).  It is easy to verify that the $s_n$ are complex and appear in complex conjugate pairs so that one may choose $a_n$
to obtain a real-valued solution.  Moreover, it can be shown that $\mathrm{Re}(s_n) < 0$ for all $n$ so that all of the modes near $\phi = K^3$ are decaying.  These decaying oscillatory modes are ghost-like and (perhaps) are related
to the presence of closed string excitations near the perturbative vacuum.

We will see shortly that, quite surprisingly, this naively linearized analysis actually gives a good qualitative picture of the fully nonlinear tachyon dynamics: generic solutions roll away from the unstable maximum and undergo damped 
oscillations about the minimum.

The reader may find it puzzling that perturbation theory around different critical points of the potential yields different numbers of initial conditions.  This kind of mis-match (which is not unique to equation (\ref{eom_SFT2}))
is an artifact of the perturbation theory employed.  We discuss the resolution of this mis-match in the appendix.  Note that because of this mis-match a naive perturbative analysis may lead to misleading results concerning
the counting of initial data in higher derivative theories.

\section{Numerical Methods}
\label{numerics}

The naive perturbative analysis employed in subsection \ref{pert_subsec} is, of course, not sufficient to establish the stability of generic solutions of equation (\ref{eom_SFT2}).  Since  we are unable to
obtain nonperturbative analytical solutions of (\ref{eom_SFT2}) we must turn to numerical analysis.  In this section we describe our numerical methods.  Although our primary interest is in equation (\ref{eom_SFT2}), we
apply our methods also to the $p$-adic string equation (\ref{eom_padic2}) as a consistency check.  Our approach follows closely the formalism developed in 
\cite{mulryne} to study nonlocal cosmological models.  Ref.\ \cite{mulryne} improved significantly on previous efforts to solve nonlocal equations numerically by allowing the equations of motion to 
be solved as an initial value problem.  Since the stability of the theory is intimately tied to the structure of the initial value problem, it is only in the context of this formulation that one can sensibly address 
the crucial issue of stability.

\subsection{Partial Differential Equation Formulation}

Infinite order differential equations such as (\ref{eom_padic2}) and (\ref{eom_SFT2}) are not directly amenable to standard numerical analysis.  In order to solve these 
equations on a computer it is convenient to introduce a fictitious auxiliary direction (which we call $r$) and re-formulate the nonlocal ordinary differential equations (ODEs)  
as local partial differential equations (PDEs) in the space spanned by the coordinates $x^{+}$ and $r$.  (This is very much analogous to the ``diffusion equation'' formulation that has been
used previously to study nonlocal cosmologies numerically \cite{nl_cosmo_joukovskaya, Joukovskaya, jouk_dimd, mulryne}.)

We start by remarking that both the equations of motion (\ref{eom_padic2}) and (\ref{eom_SFT2}) can be expressed as
\begin{eqnarray}
\label{geneom}
(1+4\xi^2\alpha \partial_+) \, e^{-\alpha \partial_+} \psi(x^+) = \psi(x^+)^p \,,
\end{eqnarray}
where we are using the notation of Ref.~\cite{mulryne}. 

For the $p$-adic case, $\xi^2 = 0$, $\alpha = -\alpha' V^+ \ln p$ and $\phi = \exp[-\alpha' V^2 \ln p/2(p-1)] \psi$, and 
for the SFT case we have $\xi^2 = - 1/(8\ln K)$, $\alpha = 2 \alpha' V^{+} \ln K$, $p = 2$ and $\phi = K^3 e^{-\alpha \partial_+} \psi$.

We now introduce an auxiliary variable $r$ and define a new field such that
\begin{equation}
\Psi(x^+,r) = e^{-r \alpha \partial_+}\psi(x^+)\,.
\end{equation}
By differentiating $\Psi(x^+,r)$ with respect to $r$ we find that $\Psi(x^+,r)$ satisfies the PDE
\begin{equation}
\label{diffusionlike}
\partial_+\Psi(t,r)=-\frac{1}{\alpha} \, \frac{\partial \Psi(t,r)}{\partial r} \,,
\end{equation}
with the boundary condition
\begin{equation}
\label{Boundary}
\Psi(x^+,1) - 4\xi^2 \left[\frac{\partial \Psi(x^+,r)}{\partial r}\right]_{r=1} = \Psi(x^+,0)^p \,,
\end{equation}
which is determined from the equation of motion (\ref{geneom}) employing the PDE (\ref{diffusionlike}). 

The nonlocal system has now been formulated in a form amenable to standard numerical methods.  Once we have specified the initial data 
$\Psi(x^{+}_{\rm i},r)$ (see the next subsection) we can proceed to numerically integrate the PDE (\ref{diffusionlike}) subject to the boundary 
condition (\ref{Boundary}) on the interval $0 \leq r \leq 1$, $x^{+} > x^{+}_{\rm i}$.  At the end of the calculation the solution $\psi(x^{+})$ 
of the original nonlocal ODE (\ref{geneom}) is extracted as
\begin{equation}
\label{extract}
  \psi(x^+) = \Psi(x^+,0)\,.
\end{equation}

Note that the system (\ref{diffusionlike},\ref{Boundary}) is very similar to the diffusion-like system obtained
in \cite{mulryne}.  In fact, it is identical with $t$ replaced by $x^{+}$, $\Box$ replaced by $\partial_{+}$.  The diffusion-like system obtained in \cite{mulryne} was ill-posed in the sense that high frequency initial 
data grew faster than lower frequency data, and hence numerical errors (which can be thought of as very high frequency noise) 
rapidly grew to swamp the real solution.  Dealing with this numerical problem was a major part of the work in \cite{mulryne}.  
The ill-posedness arises because the PDE that was solved in \cite{mulryne} is second order in time and first order in the 
auxiliary direction. In the case at hand, however, the PDE (\ref{diffusionlike}) is only first order in light-cone time and 
as a result solving this PDE is the manner described is a well posed problem.  This means that our numerical method is very stable, and the solutions produced are highly robust.

\subsection{Constructing Suitable Initial Data}
\label{bc_solve_subsec}

It remains to specify the initial data $\Psi(x^{+}_{\rm i},r)$ which must be consistent with the boundary condition (\ref{Boundary}).  In general, choosing
such initial data is nontrivial.  In order to proceed we construct suitable initial data $\Psi(x^{+}_{\rm i},r)$ perturbatively for $\psi(x^{+}_{\rm i})$
close to some value $A$.  There is no loss of generality because our formalism allows us to fix $A$ arbitrarily.  Once we have approximately determined
$\Psi(x^{+}_{\rm i},r)$ this initial configuration is then evolved numerically into the nonlinear regime.  Hence our solutions are indeed fully nonlinear.

For initial field value $\psi(x^{+}_{\rm i})$ close to some (arbitrary) constant $A$ we define 
\begin{equation}
  \psi(x^+) = A + \delta \psi(x^+)\,.
\end{equation}
Linearizing (\ref{geneom}) in $\delta\psi(x^{+})$ leads to 
%
\begin{equation}
(1+4\xi^2\alpha\partial_+) \, e^{-\alpha \partial_+} \delta\psi  = p A^{p-1} \delta\psi + A^p - A \,.
\end{equation}
%
Now $\Psi(x^{+},r) = A + \delta\Psi(x^{+},r)$ where $\delta \Psi(x^{+},r) \equiv e^{-r \alpha \partial_+} \delta\psi(x^+)$ satisfies the PDE
%
\begin{equation}
\label{PDElinear}
\partial_+ \delta \Psi = -\frac{1}{\alpha} \frac{\partial }{\partial r} \delta \Psi \,.
\end{equation}
To linear order in $\delta\psi$ the boundary condition (\ref{Boundary}) becomes
\begin{equation}
\label{Boundarylinear}
\delta\Psi(x^+,1) - 4 \xi^2 \left[\frac{\partial \delta\Psi(x^+,r)}{\partial r}\right]_{r = 1} =
 p A^{p-1} \delta\Psi(x^{+},0) + A^p -A \,.
\end{equation}

We can now solve for $\delta\Psi(x^+,r)$ by separation of variables.  We take an ansatz of the form
\begin{equation}
\label{SepSol}
\delta\Psi(x^+,r) = \delta\psi(x^+) g(r) + h(r) \,.
\end{equation}
Substituting (\ref{SepSol}) back into (\ref{PDElinear}) we find that the functions dependent on the auxiliary variable $r$ 
are of the form $g = e^{-\alpha\omega^2 r}$ and $h = b (e^{\alpha \omega^2 r}-1)/\omega^2$, while $\delta\psi$ satisfies the local equation
\begin{equation}
\label{LocalField}
\partial_+ \delta\psi = -\omega^2 \delta\psi -b \,.
\end{equation}
Here $\omega^2$ is any solution of the characteristic equation
\begin{equation}
\label{characteristic1}
e^{\alpha\omega^2}(1-4\xi^2\alpha\omega^2) = pA^{p-1} \,,
\end{equation}
and $b$ is given by
\begin{equation}
\label{b1}
\frac{b}{\omega^2} = \frac{A^p-A}{pA^{p-1}-1} \,.
\end{equation}

In general the characteristic equation (\ref{characteristic1}) will have many roots $\omega_n^2$. 
For each root $\omega_n^2$ of the characteristic equation we can obtain a solution $\delta\psi_n(x^{+})$ of equation (\ref{LocalField})
which, physically, can be thought of as a particle-like excitation near $\psi = A$.  The spectrum of roots $\omega_n^2$ corresponds 
to the spectrum of masses for these physical excitations.  
For each state $\delta\psi_n(x^{+})$ there also exists a particular solution of the
PDE: $\delta\Psi_n(x^{+},r) = e^{\alpha \omega_n^2 r} \delta\psi_n(x^{+}) + b_n (e^{\alpha \omega_n^2 r}-1 )/\omega_n^2$.  
To construct general solutions, of course, we must superpose these modes, and we are 
lead to the general solution
\begin{equation}
\label{GenSol}
\delta \Psi(x^+,r) = \sum_n \left[ \delta \psi_n(x^+) e^{\alpha\omega_n^2 r} + \frac{b_n}{\omega_n^2}\left(e^{\alpha\omega_n^2 r}-1\right) \right] \,,
\end{equation}
where it can be easily verified that the constants $b_n$ must now satisfy the relation 
\begin{equation}
\sum_n \frac{b_n}{\omega^2_n}=\frac{A^p-A}{pA^{p-1}-1}\,
\end{equation}
which replaces (\ref{b1}) when we superpose more than one mode, but are otherwise arbitrary.

\begin{figure}
\begin{center}
\includegraphics[width=8cm]{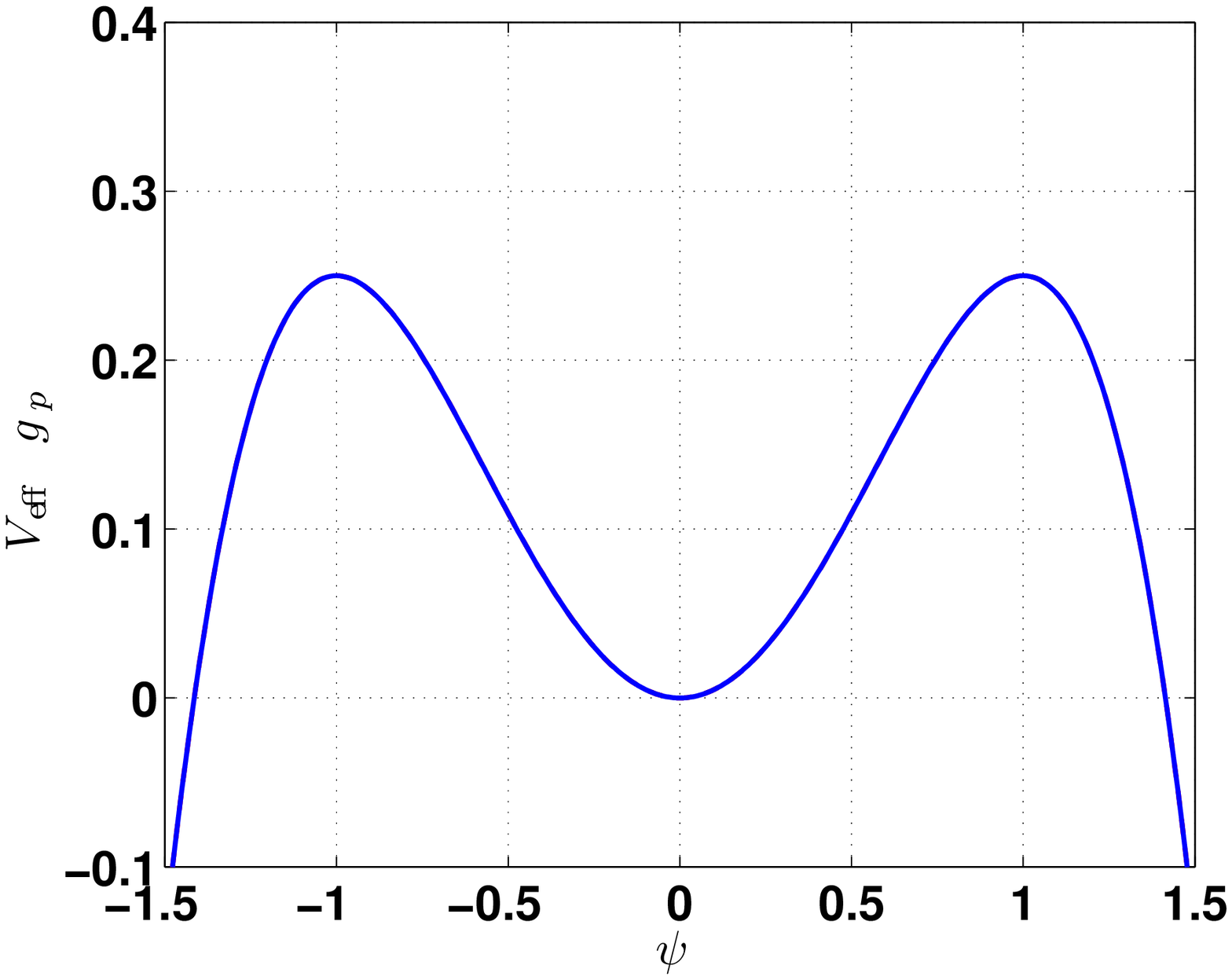} 
\caption{\label{veff1} The $p$-adic string effective potential with $p = 3$.}
\end{center}
\end{figure}

\begin{figure}
\begin{center}
\includegraphics[width=7cm]{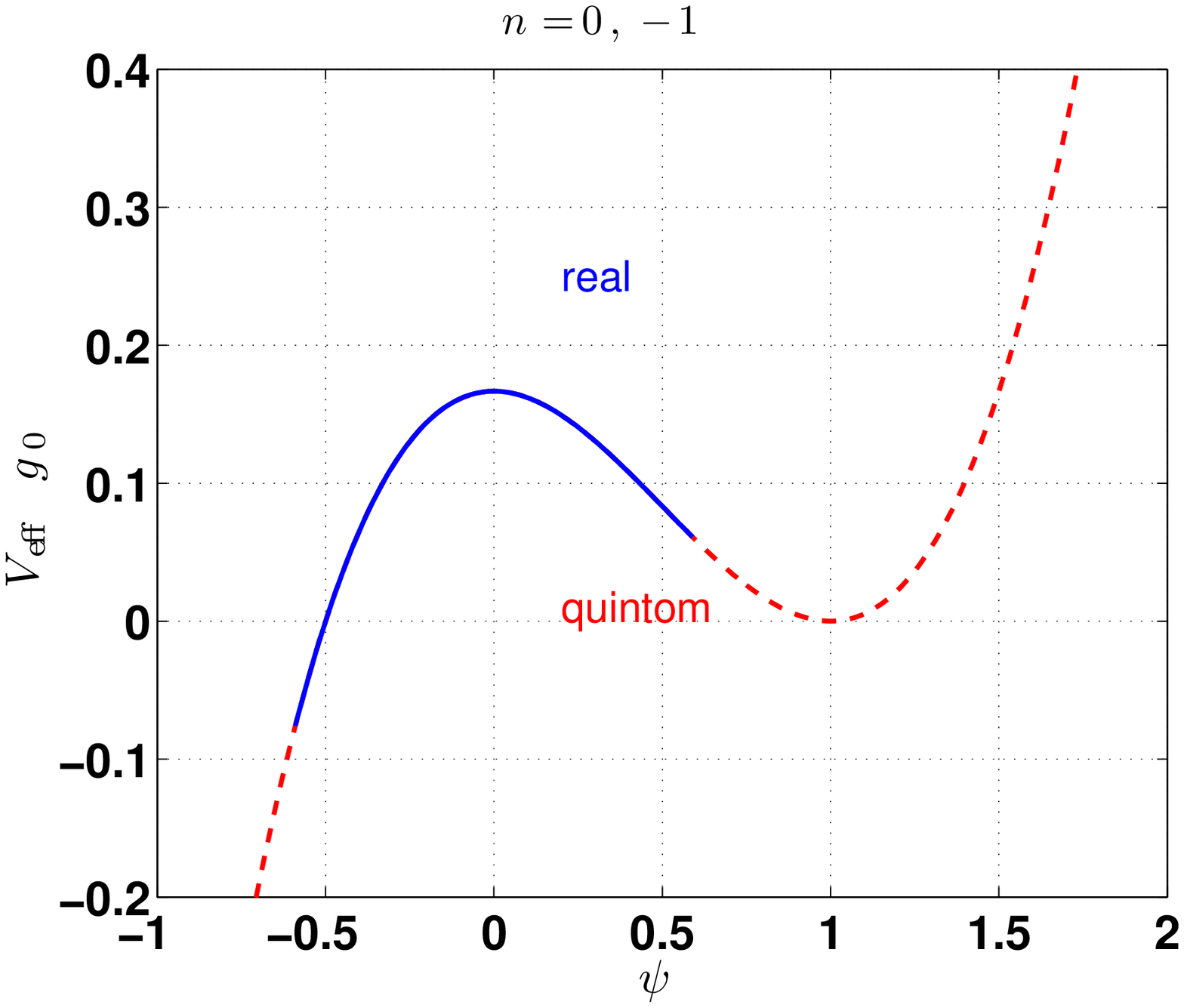}
\includegraphics[width=7cm]{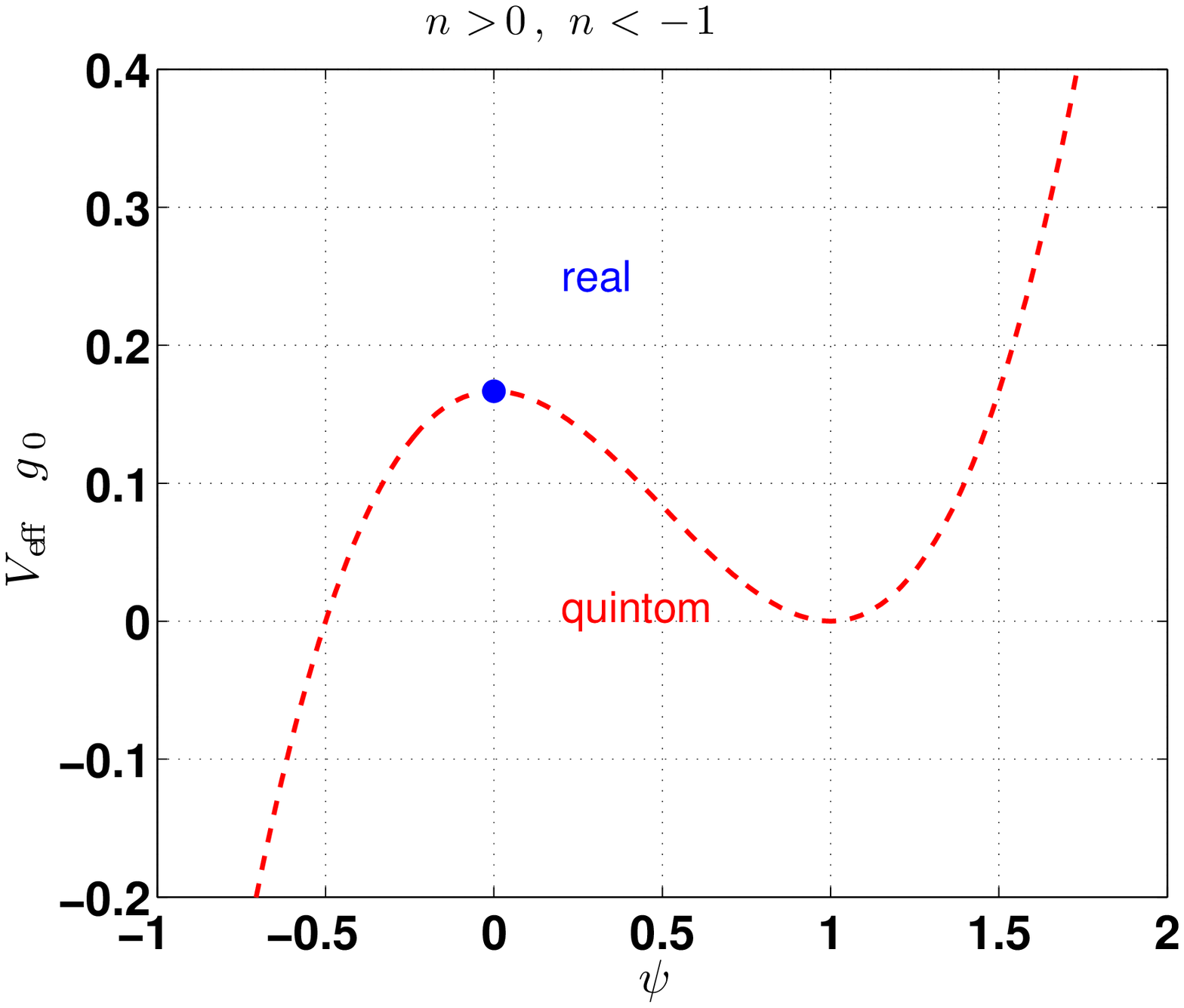}
\caption{\label{veff2} The effective potential of the SFT case. The left panel shows the regions where the dynamics is initially described by a real field (solid line) or as a quintom (dashed lines) for $n = 0, -1$.
 The right panel shows the same regions for the remaining values of $n$.}
\end{center}
\end{figure}

When $\xi^2 = 0$, such as in the $p$-adic case, the roots
 are given by
\begin{equation}
\label{roots1}
\alpha \omega_n^2 = \ln (pA^{p-1}) \pm i 2\pi n \,, 
\end{equation}
and when $\xi^2 \ne 0$, like in the SFT case, we obtain 
\begin{equation}
\label{roots2}
\alpha\omega_n^2 = \frac{1}{4\xi^2} + W_n\left[-\frac{pA^{p-1}}{4\xi^2} e^{-1/4\xi^2}\right]\,, 
\end{equation}
when $A\neq0$ and 
\begin{equation}
\alpha\omega^2 = \frac{1}{4\xi^2}  \,,
\end{equation}
when $A = 0$.  $W_n$ again represents the branches of the Lambert-W function. For both the $p$-adic and SFT cases, if $A$ is sufficiently close to the unstable maximum then there exists a real-mass mode with 
$\alpha\omega^2 > 0$.  This mode corresponds to the usual tachyon and reflects the instability
of the false vacuum.  (There is also a decaying mode with real-mass which drops out of the spectrum very quickly.)  In addition to this tachyonic state we also have an infinite tower of states with complex mass-squared.  The kinetic 
energy associated with these states has indefinite sign \cite{pais} and hence 
these states are ghost-like.  More precisely, each complex-mass state behaves as an admixture of a ghost field and a non-ghost field \cite{mulryne} (hence we use ``ghost-like'' rather than ``ghost'').  In \cite{mulryne} 
such states were described as ``quintoms'' and we will occasionally use this term interchangeably with ghost-like.
The dependence of the mass spectrum on the initial value $A$ is illustrated in Figs.~(\ref{veff1}) and (\ref{veff2}).  There we show the effective potential of the models under study and the regions where the characteristic 
equation has at least one real root and regions where all the roots are complex (which can be compared to the time-like cosmological cases in Ref.~\cite{proceedings}). 


We can see from the preceding discussion and from Figs.~(\ref{veff1}) and (\ref{veff2}),  that 
the characteristic equation, for a given point $A$, has at most two real roots and an infinite number of complex roots. 
Hence, a particular solution $\delta \Psi_n(x^+,r)$ is, in general, complex-valued.  For each complex root, 
its complex conjugate is also a root, hence, $\delta \Psi^*_n$ is also a particular solution 
to the PDE equation.  Moreover, by suitably combining $\delta\Psi_n$ and $\delta\Psi^*_n$, we can construct real-valued solutions. 
Having this in mind, we are now ready to specify suitable initial data for our simulations. We are particularly interested in the evolution which follows when the field is initially in a region close to the maximum of the effective potential. It can readily be verified that an acceptable initial profile in the $p$-adic case, where the maximum exists at $\psi = 1$, is
\begin{equation}
\label{IC1}
\Psi(x^+_{\rm i}) = 1-\epsilon \sum_{n = 0} a_n e^{\alpha {\omega_n}_R^2 r} \, \cos(\alpha{\omega_n}_I^2 r + \theta_n) \,,
\end{equation}
and an initial profile for the SFT case, where the maximum is at $\psi = 0$,
can be written as
\begin{equation}
\label{IC2}
\Psi(x^+_{\rm i}) = \epsilon \sum_{n=0} a_n e^{\alpha{\omega_n}_R^2 r} \, \cos(\alpha{\omega_n}_I^2 r + \theta_n) \,,
\end{equation}
where $\theta_n$ is an arbitrary phase, the indices $R$ and $I$ mean the real and imaginary parts, and $\epsilon$ is a small number.

In equations (\ref{IC1}), (\ref{IC2}) the free coefficients $a_n$ allow us to fix an infinite number of initial conditions.  Physically each $a_n$ parameterizes the amount of the state with
mass-squared $\omega^2_n$ that is present in the initial admixture.  The approximate initial functions (\ref{IC1}), (\ref{IC2})  can now be numerically evolved using (\ref{diffusionlike}), (\ref{Boundary}) into 
the fully nonlinear regime.


For simplicity, we have set the arbitrary phases $\theta_n$ to zero in our examples.
We have verified that the inclusion of $\theta_n \not= 0$ does not qualitatively change
the behaviour of our solutions.  In particular, the inclusion of these phases does not change our results concerning the nonlinear stability of generic SFT solutions.

\subsection{Consistency Check Using the $p$-adic Theory}

Although our motivation for turning to numerical analysis was equation (\ref{eom_SFT2}), we can also use this method as a consistency check on our previous results for the $p$-adic string (section \ref{light_padic}).
First, let us verify that we can reproduce the analytic solution (\ref{phi}) using the PDE formulation.  To this end, we introduce another field $\Upsilon(x^{+},r)$ related to $\Psi(x^{+},r)$ by $\Upsilon = \ln \Psi$.  In terms of this
new field the boundary condition (\ref{Boundary}) is linear: $\Upsilon(x^+,1) = p \Upsilon(x^+,0)$.  Now we can solve the diffusion-like equation (\ref{diffusionlike}) exactly by separation of variables, 
$\Upsilon = f(x^+) g(r)$ where $f(x^+)$ and $g(r)$ are of the form
\begin{eqnarray}
f(x^+) &=& e^{-\omega^2 x^+} \,, \\
g(r) &=& e^{\alpha\omega^2 r} \,,
\end{eqnarray}
The roots of the characteristic equation $\alpha\omega^2_n$ are determined using the boundary conditions to be:
\begin{equation}
\alpha \omega^2_n = \ln p \pm i2\pi n \,.
\end{equation}
Putting all these results together and using (\ref{extract}) leads to the same solution for $\phi(x^+)$ found in Eq.~(\ref{phi}). This 
confirms that the PDE method is consistent with other methods of solving non-local equations of motion.

Note that for the SFT case, the boundary condition (\ref{Boundary}) cannot be  written in terms of a linear relation, and therefore the method of separation of 
variables cannot be employed to obtain a solution to the PDE. This is simply a reflection of the statement we made previously that we cannot find an easy analytic 
solution to Eq. (\ref{eom_SFT2}) (unlike the $p$-adic case Eq. (\ref{eom_padic})).

\begin{figure}
\begin{center}
\includegraphics[width = 8cm]{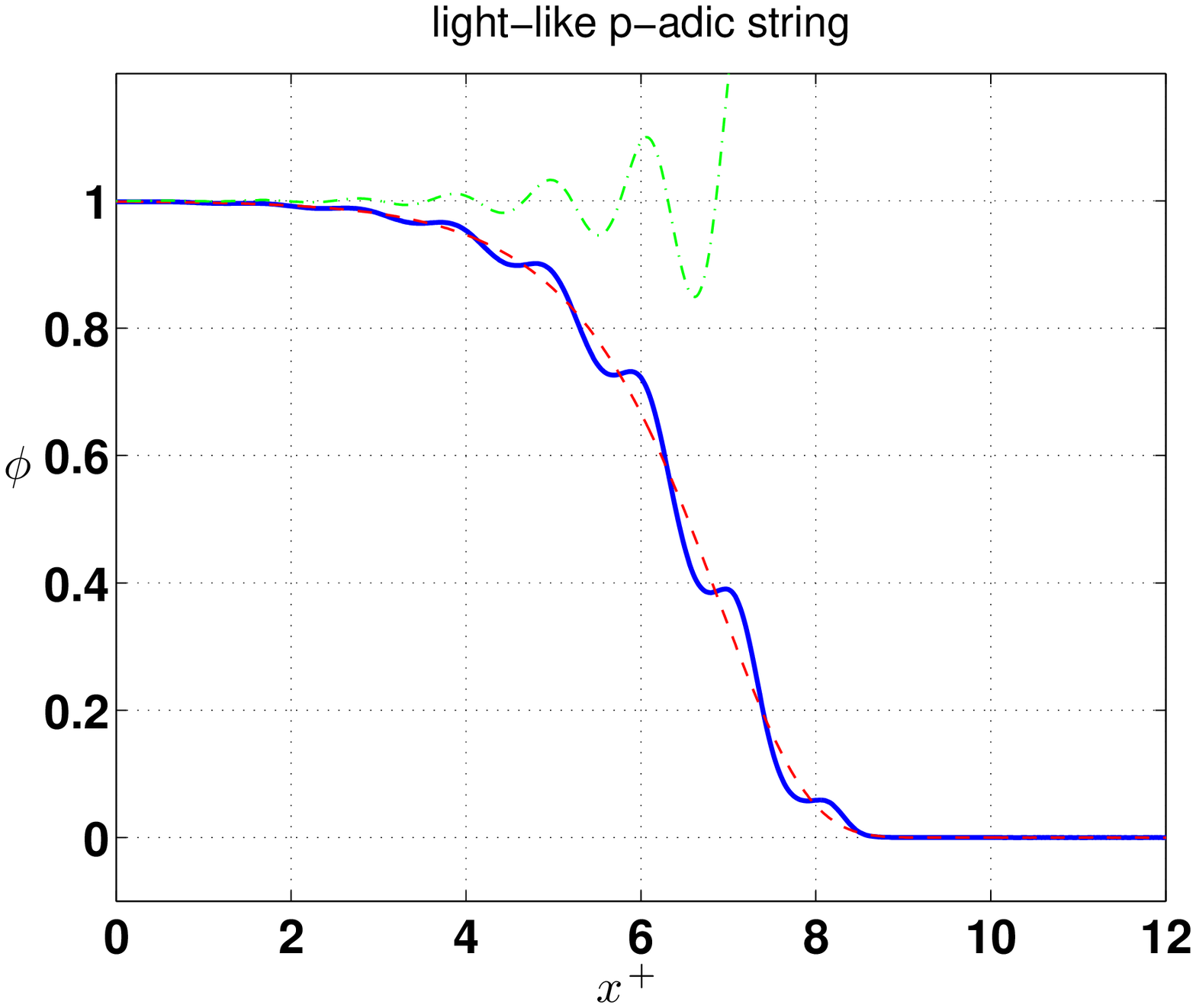}
\caption{\label{fig:llpadic} Evolution of the combination of the real field with the first quintom ($n = 1$)  in the $p$-adic string (solid line). The dashed and dash-dotted lines represent the evolution of the real field and of the quintom, respectively, when they are  rolling in isolation. We have used $\epsilon = 10^{-3}$, $a_0 = 1$, $a_1 = 0.25$, $p = 3$, and $V^2 = 0$. The $x^{+}$ axis is measured in units of $\alpha'V^+$. }
\end{center}
\end{figure}

As a final consistency check we compare numerical solutions of (\ref{diffusionlike}) and (\ref{Boundary}) using the approach of subsection \ref{bc_solve_subsec} to the exact analytic solutions obtained in section \ref{light_padic}.
In Fig.~\ref{fig:llpadic} we plot the light-cone time evolution of the $p$-adic tachyon for a combination of the real field ($n = 0$) with a oscillatory mode, which we refer to as a quintom in line with \cite{mulryne}. 
We choose the $n = 1$ mode. We see that the real field helps the quintom to decay and to remain harmless for the entire length of the subsequent evolution, assuming that the amount of quintom present initially does not 
violate the condition (\ref{cond1}) which we derived analytically above.  Moreover, the numerical solution is in excellent agreement with the analytical solution (\ref{phi}), a 
further verification of both the interesting behaviour of that solution, and of our numerical method.

\section{Nonlinear Stability of the Light-Like SFT Tachyon}
\label{SFT_num_sec}

\subsection{Numerical Solutions}

Having described in detail our numerical methods in section \ref{numerics} we now wish to apply these to study the nonlinear dynamics of equation light-like tachyon condensation in SFT for arbitrary initial data.
We proceed by numerically integrating the 
PDE (\ref{diffusionlike}) using the non-linear boundary conditions (\ref{Boundary}), from initial data of the form (\ref{IC1})--(\ref{IC2}) determined 
by perturbing about the hill-top as we have just described above. We stress that, although the initial data is fixed by considering a linearization of the equations of motion, the numerical solution rapidly evolves out of this linear 
regime and behaves in a fully non-linear manner.

\begin{figure}
\begin{center}
\includegraphics[width = 8cm]{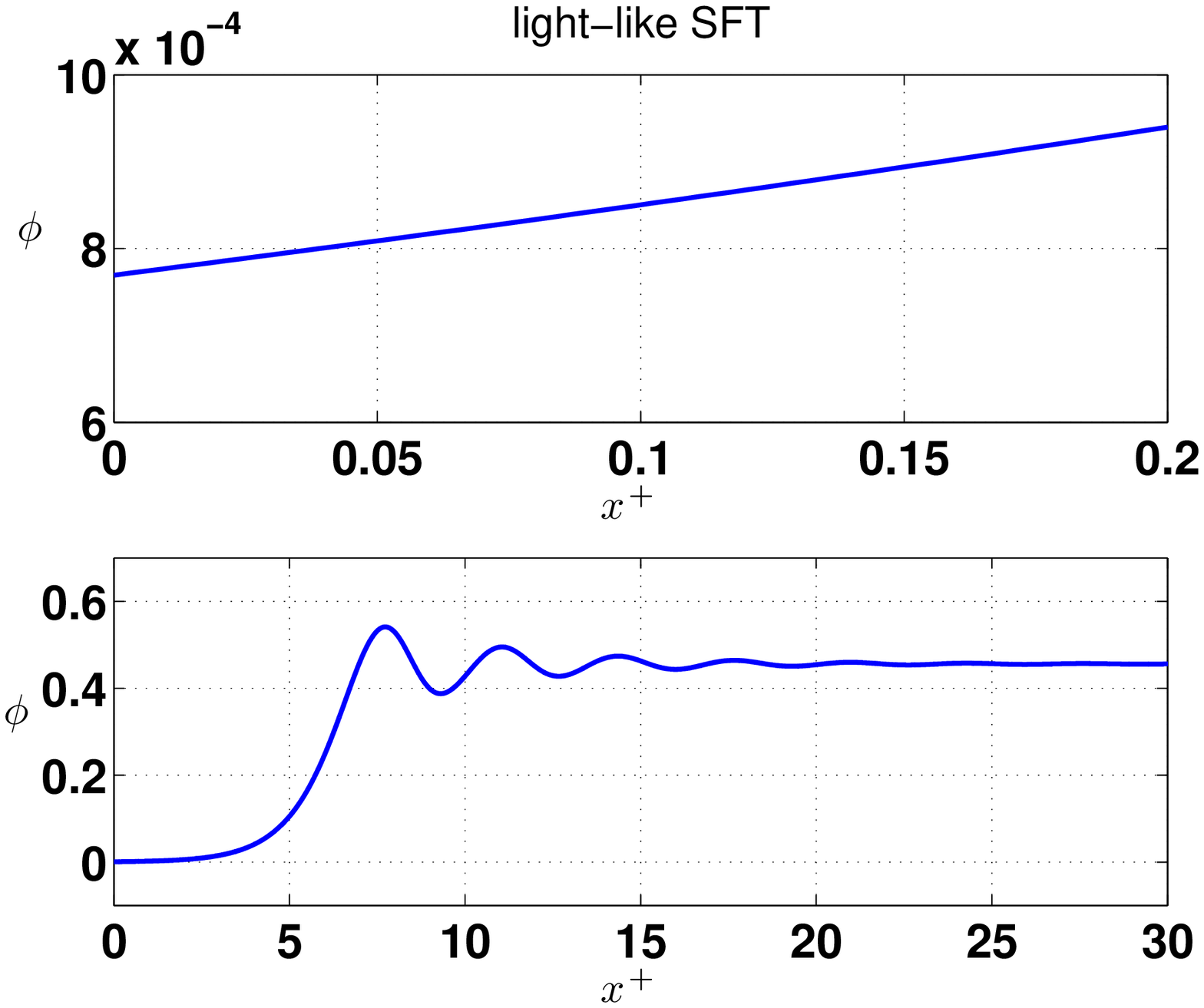}
\caption{\label{fig:llsft1} Evolution of the real field $n = 0$ in the SFT case. The upper panel shows the early evolution where we can see that the field rolls straight to the minimum of the potential. The lower panel shows the full evolution.}
\end{center}
\end{figure}

First, we verify that picking initial conditions such that the real field tachyon is 
present initially, leads to the well behaved rolling solution which 
was constructed in \cite{hs}. Figure \ref{fig:llsft1} presents the 
rolling solution produced by our numerical method 
which follows from picking only the $n=0$ initial condition. As can be seen, this solution has exactly 
the behaviour expected.  

\begin{figure}
\begin{center}
\includegraphics[width = 8cm]{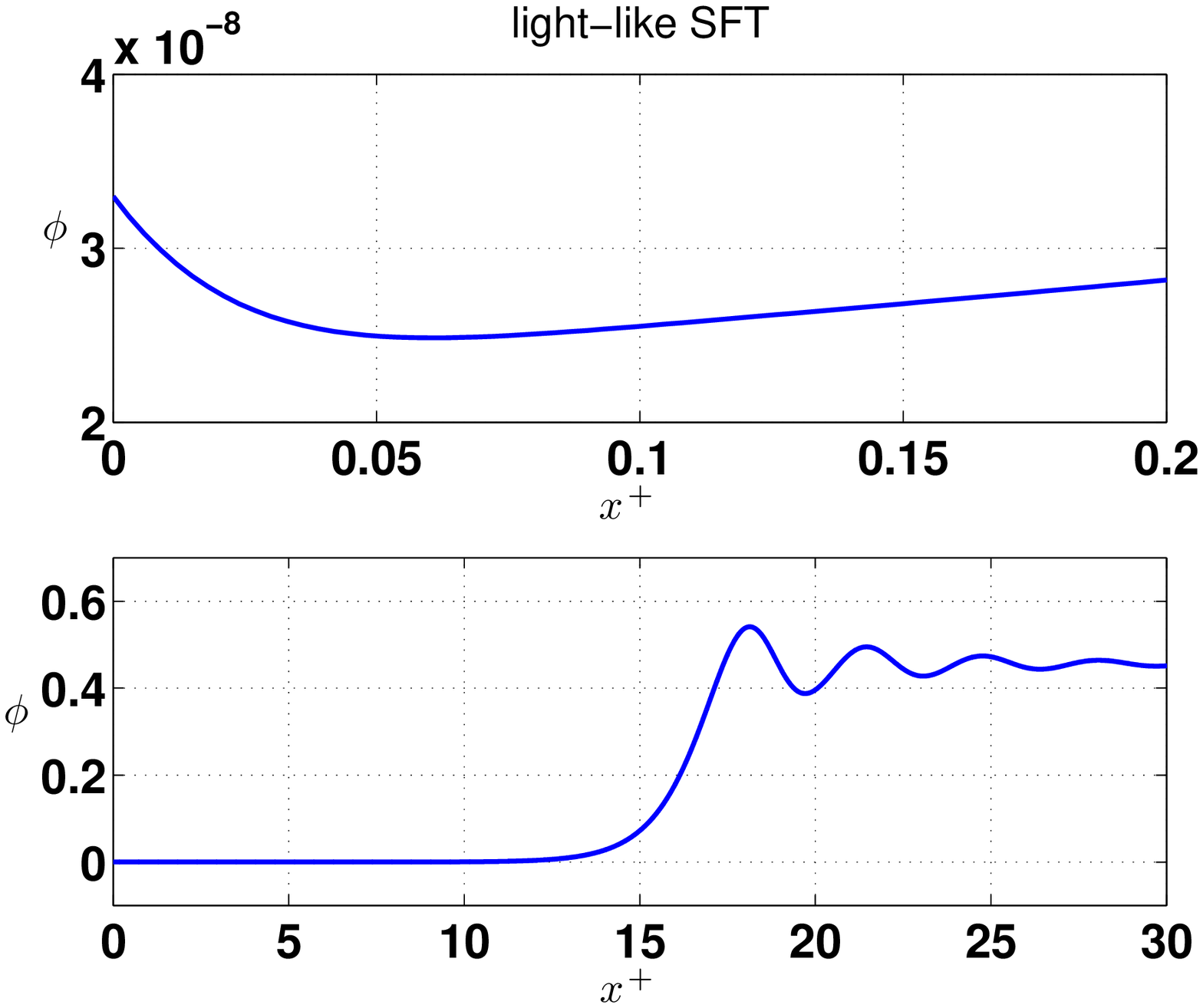}
\caption{\label{fig:llsft} Evolution of the quintom $n = 1$ in the SFT case. The upper panel shows the early evolution where we can see the quintom decaying towards the maximum of the potential. The lower panel shows the full evolution where it is clear that only a real field evolving towards the minimum subsists.}
\end{center}
\end{figure}

We now focus on the fate of the ghost-like excitations.  
In Fig.~\ref{fig:llsft} we show that the quintoms decay on their own in the SFT case without need 
of the extra contribution from the real field. Starting the evolution with only the $n=1$ quintom present, 
we see that this excitation 
first decays towards the unstable maximum, and then at late times a real field 
is formed which evolves towards the minimum of the effective potential performing damped oscillations (assuming, of course,  that the real field does not start to roll down the unbounded side of the potential). 
The initial decay is in line with the expectations drawn from a perturbative analysis close to the 
hill-top, then the subsequent roll to the minimum  is the natural consequence of the maximum being an unstable point which the real field tachyon 
naturally tries to roll away from. 

We have explored a wide range of other initial conditions for the SFT case, and other choices of $n$, and the situation is always the same: any ghost-like excitation present initially decays, as we expected from a perturbative 
analysis, and at late times the rolling solution is reached and the field decays to the minimum of the potential. Indeed even if we fix our initial conditions near to the minimum of the potential by perturbing about a point near 
to the minimum (where the roots to the characteristic equation are all complex and we can only have ghost-like states present as an initial condition), the behaviour that is always found is a rapid decay to the minimum.

\subsection{Ghosts, Stability and Friction}

The dynamics of equation (\ref{eom_SFT2}) are quite remarkable.  This equation admits infinitely many initial conditions and, by Ostrogradksi's theorem, the Hamiltonian is unbounded from below.  However, the dilaton gradient
violates time translation invariance so that intuition with Hamiltonian dynamics gives a completely inaccurate picture of the actual behaviour of the tachyon.  Before getting too excited, we should remember that this stability
relies crucially on the ansatz $\phi = \phi(x^{+})$ and that on small scales (compared to the dilaton gradient) generic field profiles should display the usual unstable behaviour.  With this caveat in mind, however, we believe that 
our findings our quite significant.  Equation (\ref{eom_SFT2}) provides an explicit example of an interacting nonlocal theory that admits infinitely many initial conditions but is completely stable.  It is our hope that this toy example will 
provide hints into how to construct more realistic stable theories with infinitely many derivatives. 

It is worth noting that our discussion of stability here refers only to the kinds of erratic time evolution that are associated with the Ostrogradski instability.  Of course, the potential for the theory (\ref{eom_SFT2}) is unbounded from
below and hence there is an instability associated with rolling down that unbounded direction.  This instability would be present also in a local field theory with the same potential and is completely unrelated to the kinds of higher
derivative instabilities that we are interested in.  The unboundedness of the potential is thought to be physically associated with the closed string tachyon.

\section{Stability of Nonlocal Cosmologies}
\label{cosmo_sec}

Owing to the close similarity between the equations describing light-like tachyon condensation and the equations describing nonlocal cosmologies (such as $p$-adic inflation) one may wonder whether the kinds of phenomena discussed above
occur also in the latter case.  It has been observed previously that it is \emph{possible} to obtain cosmological solutions where the tachyon settles down to the true vacuum at late times \cite{jouk_dimd, mulryne}.  However, previous 
studies have not considered how generic such solutions are, and indeed other wildly oscillatory solutions have also 
been produced \cite{mulryne}.  

We have numerically studied both $p$-adic and SFT cosmologies for a variety of initial conditions describing a mixture of tachyon and ghost-like excitation in the initial state.  Our preliminary results suggest that, at least for 
certain parameter choices, the set of initial conditions leading to 
non-pathological evolution is not measure zero.  Hence, we are lead to suspect that the island of stability is a fairly general property of infinite order theories in the presence of friction.  This suggests that the combination of 
friction and constraints on the initial data provides a very general recipe for obtaining stable solutions in higher derivative theories.\footnote{Note that these kinds of constraints could be elegantly implemented at the level of the 
formal pseudo-differential operator by using the contour deformation prescription advocated in \cite{niky}.}  This observation should be helpful to guide future attempts to construct physically sensible time-dependent solutions in SFT.

Before we leave this section a few comments are in order.  As discussed previously, in the cosmological context our numerical methods are much less robust than in the light-like case.  Although our preliminary efforts suggest
that the island of stability exists in the cosmological context, it is difficult to make conclusive statements because our finite numerical accuracy prevents us from following the evolution to arbitrarily late times.  Hence, although
our solutions \emph{appear} to settle down to the minimum of the potential, we cannot (yet) rule out the possibility that instabilities re-appear at very late times.  (Note that this problem does \emph{not} afflict our light-like
solutions discussed previously.)  We intend to return to this issue in future work.

\section{Are Wild Oscillations Necessarily Catastrophic?}
\label{wild_sec}

Throughout this paper we have discussed the nonlinear dynamics of the infinite order equations that describe tachyon dynamics in string theory.  Our focus has been on constructing solutions that are not afflicted by
wild oscillatory (or otherwise unstable) behaviour at late times.  However, we have not considered the question of whether such instabilities are necessarily catastrophic.  If a field theory such as (\ref{Spadic}) or
(\ref{Ssft}) were the complete picture the answer might be straightforward, however, the problem is rather more subtle in the context of string field theory.  The stress tensor at a point is not a truly gauge invariant
object since neither the interactions of off-shell close string vertex operators, nor the restriction of the boundary state zero mode to particular values, preserves BRST invariance \cite{hs}.  Since it is not trivial to
relate the tachyon $\phi$ in the level truncation to physical observables it is not entirely clear if the wild oscillations are problematic.  Several different explanations have been proposed in the SFT literature.  We briefly
discuss these below.

Erler and Gross argued, using a clever choice of basis (the light-cone basis), that the full SFT is first order in a single null direction (say $x^{+}$) and nonlocal in all the remaining directions \cite{gross}.  The initial value formulation in this case
is somewhat more complicated than in nondegenerate second order systems (where one specifies the coordinates and velocities at $t=0$).  In the Erler and Gross formulation one must provide information about the field at $x^{+} = 0$
and also on the surface $x^{-} = c$ for all $x^{+}$ (here $x^{-}$ is the other null direction orthogonal to $x^{+}$).  In \cite{gross} it was argued that one can take $c \rightarrow -\infty$ and demand that the field vanish in this 
limit.  Although this requirement is physically sensible, it is not clear if it represents an undue restriction on the free coefficients of the solution.  Note also that practical computations in the light-cone basis are complicated 
by the re-appearance of spurious negative energy states as artifacts of the level truncation \cite{erler}.  

Coletti et al.\ \cite{taming} argued that the wild oscillations of the tachyon field can be eliminated by a nonlocal transformation of the form $\phi(t) = f(\partial_t)T(t) + \cdots$ which takes the cubic open string field theory 
action to the analogous boundary string field theory action.  The generatrix $f(s)$ in this case has both poles and zeroes in the complex $s$-plane and hence one might worry about changing the number of degrees of freedom in the solution 
(see, for example, \cite{niky}).

Finally, Kiermaier et al.\  \cite{boundary} constructed a class of BRST-invariant closed string states for any classical solution of open string field theory.  This state can be used to provide gauge-invariant observables.  
Using this state Kiermaier et al.\ argue that the wildly oscillatory rolling tachyon solution actually describes the regular close string physics studied in \cite{sen_rev}.  The peculiar time evolution is interpreted
as resulting because the regular physics of the closed string sector is being described in terms of open string degrees of freedom.

It is sometimes argued that the ghost-like modes present in level-truncated SFT and $p$-adic string theory are artefacts.  In this case one expects that more realistic dynamics will be obtained by projecting these states out, presumably
through some prescription for choosing initial data (see \cite{niky} for an elegant implementation).  Our analysis of the island of stability can be thought of as elucidating the minimal constraint on the initial data necessary to obtain
sensible evolution.  The idea of projecting out ghost excitations from higher derivative theories using some boundary conditions is not new.  Qualitatively similar prescriptions have been employed in the finite derivative case; 
see \cite{hh} and \cite{LW}-\cite{LW-SM}.

\section{Conclusions}
\label{concl_sec}

Using a combination of analytical and numerical methods we have investigated the nonperturbative stability of light-like rolling tachyon solutions in the presence of a linear dilaton background.  
We have uncovered some potentially surprising results.  We have seen that the addition of friction can drastically soften the effects of higher derivative instabilities.  In the case of 
the $p$-adic string (and also VSFT) we have found an island of stability in initial condition space.  For initial conditions within this island the tachyon dynamics are non-pathological.  Interestingly, the 
island of stability is not a set of measure zero nor is it an artifact of perturbation theory.  We have found qualitatively similar behaviour in the cosmological context and have speculated that the recipe of
mixing friction with some constraints on the initial data provides a general prescription for constructing sensible (particular) solutions in nonlocal theories. 

In the case of SFT at level zero truncation the effect of the friction on the higher derivative instability is even more dramatic.  In this case the unstable growth associated with the ghost-like modes is completely damped out by the dilaton
gradient and the resulting tachyon dynamics are non-pathological for generic choice of initial conditions!  This provides an invaluable example of an interacting nonlocal theory derived from string theory that is completely stable.
A caveat is that this stability relies on the ansatz $\phi=\phi(x^+)$ and, as we have argued, is not expected to persist in the case where $\phi$ depends on all space-time coordinates.  We do not believe that this caveat should be viewed as a
serious limitation on our analysis since the significance of our results does not lie in the claim that light-like profiles are the most realistic rolling tachyon solutions.  Rather, our results are important because we have uncovered a
previously unexpected loop-hole in Ostrogradski's theorem. 

During the course of our investigation we have developed many general techniques for studying nonlocal theories at the fully nonlinear level.  It is our hope that
these results/techniques will lead to the discovery of more realistic examples of stable infinite order theories.

\section*{Acknowledgments}

This work was supported in part by NSERC. DJM is supported by the Centre for Theoretical Cosmology, Cambridge, NJN is supported by Deutsche Forschungsgemeinschaft, TRR33 and PR 
is supported by an NSERC USRA. We are grateful to T.\ Biswas, J.\ Cline, S.\ Hellerman, N.\ Kamran, M.\ Schnabl and R.\ Woodard for helpful discussions and correspondence.

\renewcommand{\theequation}{A-\arabic{equation}}
\setcounter{equation}{0}  

\section*{APPENDIX: The Perturbative Mis-Match}

In subsection \ref{pert_subsec} we discussed the perturbative construction of solutions of equation (\ref{eom_SFT2}) about the constant solutions $\phi = 0, K^3$.  There we were presented with the 
curious puzzle that  - taking this result seriously - one would infer different numbers of initial conditions about these two critical points.  This mis-match is an artifact of the perturbation
theory that was employed.  To see this, we reconsider solving equation (\ref{eom_SFT}) in a linearized expansion about an arbitrary value $\phi = \phi_0$.  We could simply extract the result from
our analysis in subsection \ref{bc_solve_subsec}.  However, the formlism does not rely on the PDE formulation and it may be of interest to show how to perturb about an arbitrary $\phi_0$ 
without introducing the auxiliary direction $r$.  

We begin in a very general context and specialize to equation (\ref{eom_SFT2}) at the end.  Consider the nonlocal equation
\begin{equation}
\label{gen}
  F(D) \phi = V'\left[\phi\right]
\end{equation}
where $F(z)$ is an analytic function of the complex variable $z$ that can be represented by a convergent series expansion
\begin{equation}
\label{F(z)}
  F(z) = \sum_{n=0}^{\infty} a_n z^n
\end{equation}
and $D$ is some linear differential operator which satisfies $D (\mathrm{const}) = 0$.  We wish to solve (\ref{gen}) near some 
constant value $\phi = \phi_0$ not necessarily a solution of (\ref{gen}).  Writing 
\begin{equation}
  \phi = \phi_0 + \delta \phi
\end{equation}
and linearizing in $\delta \phi$ we have
\begin{equation}
\label{gen_pert}
  \left[ F(D) - V''\left[\phi_0\right] \right] \delta \phi + \left[F(0) \phi_0 - V'\left[\phi_0\right] \right] = 0
\end{equation}
If $\phi_0$ were a solution of (\ref{gen}) then the second term in the square braces would vanish.  In this case one could construct
the solution $\delta\phi$ by provisionally taking $\delta \phi$ to be an eigenfunction of $D$ (that is assuming $D \delta \phi = -\omega^2\delta \phi$).  
This is precisely the approach that was adopted in \cite{pi}-\cite{ng2} and \cite{niky2}.  However, we can easily generalize this approach more general
values of $\phi_0$ by provisionally taking $\delta \phi$ to be a solution of the equation
\begin{equation}
\label{gen_ansatz}
  D \delta \phi = -\omega^2 \delta \phi - b
\end{equation}
It is straightforward to show that the action of the pseudo-differential operator $F(D)$ on a function satisfying (\ref{gen_ansatz}) is
\begin{equation}
\label{gen_identity}
  F(D) \delta \phi = F(-\omega^2) \delta \phi + \frac{b}{\omega^2}\left[ F(-\omega^2) - F(0) \right]
\end{equation}
Plugging (\ref{gen_identity}) into (\ref{gen_pert}) we find that the solutions of (\ref{gen_ansatz}) are also solutions of the fully nonlocal equation (\ref{gen_pert})
as long as $\omega$, $b$ are chosen to satisfy the following algebraic equations:
\begin{eqnarray}
  0 &=& F(-\omega^2) - V''\left[\phi_0\right] \label{roots} \\
  b &=& \frac{-\omega^2}{V''\left[\phi_0\right] - F(0)}\left[  F(0) \phi_0   -  V'\left[\phi_0\right] \right] \label{b}
\end{eqnarray}
In general equation (\ref{roots}) may have multiple roots $\omega_n$ ($n = 1, \cdots, N$) which defines $N$ solutions $b_n$ of equation (\ref{b}).  There are then $N$ independent solutions 
$\delta \phi_n$ (each solving an equation of the form $D\delta\phi_n = -\omega_n^2\delta\phi_n -b_n$) and the most general solution of (\ref{gen_pert}) is obtained by superposing these modes as
\begin{equation}
  \delta \phi = \sum_{n=1}^N \delta \phi_n
\end{equation}
When superposing multiple roots equation (\ref{b}) generalizes to
\begin{equation}
\label{bn_app}
  \sum_n \frac{b_n}{\omega_n^2} = \frac{- F(0) \phi_0   +  V'\left[\phi_0\right]}{V''\left[\phi_0\right] - F(0)}
\end{equation}
Note that this method may fail if $\omega^2 = 0$ is a solution of (\ref{roots}).  This approach is identical to the formalism employed in \cite{mulryne} and also in subsection \ref{bc_solve_subsec}.

Now we specialize to equation (\ref{eom_SFT2}).  We take $D = \partial_{+}$, $F(z) = (\alpha'V^{+} z - 1)K^{-2\alpha'V^{+}z}$ and $V'\left[\phi\right] = -\phi^2 / K^3$.  After some straightforward manipulations
we find that the mode functions $\delta\phi_n$ take the form
\begin{equation}
  \delta\phi_n(x^{+}) = a_n e^{s_n x^{+}} + \frac{b_n}{s_n}
\end{equation}
where the $a_n$ are arbitrary constants and
\begin{equation}
\label{gen_sn}
  \alpha' V^{+} s_n =  1 - \frac{1}{2\ln K}W_n\left[ \frac{4\ln K}{K} \phi_0 \right] 
\end{equation}
(where $W_n$ denotes the branches of the Lambert-W function and $n$ runs over all integer values).  Summing over the modes and making use of equation (\ref{bn_app}) we have
\begin{eqnarray}
  \delta \phi(x^{+}) &=& \sum_n \delta \phi_n \nonumber \\
                     &=& \sum_n a_n  e^{s_n x^{+}} - \phi_0  \frac{1-\phi_0 / K^3}{1-2\phi_0 / K^3}
\end{eqnarray}
For $\phi_0 = K^3$ these expressions reproduce equations (\ref{vac_mode}) and (\ref{vac_freq}).  Let us consider
the case $\phi_0 = 0$, where a naive perturbative analysis yields only a single growing mode.  Taking the limit $\phi_0 \rightarrow 0$ in equation (\ref{gen_sn}) we find that
$s_0 \rightarrow 1/(\alpha'V^{+})$ while $\mathrm{Re}(s_n) \rightarrow -\infty$ for all $n \not= 0$.  Hence, the infinite tower of ghost-like modes decay very quickly near the false
vacuum.  In the limit $\phi_0 \rightarrow 0$ these spurious states all go to zero infinitely fast and completely drop out of the spectrum.  Thus, the analysis in this appendix is consistent
with equation (\ref{single_grow}) and the puzzle of the mis-match of initial data counting is resolved.

A similar mis-match would occur if we have studied the $p$-adic equation (\ref{eom_padic2}) in a perturbative expansion about the constant solutions $\phi = 0, p^{-\alpha' V^2/\left[2(p-1)\right]}$.
There one finds infinitely many solutions about the false vacuum and no non-trivial solutions at the true vacuum.  Again, the mis-match is an artifact as can be seen by examining the fully nonperturbative
solution (\ref{phi}).  From this solution it is clear that the extra states become nonperturbative near $\phi = 0$.

\end{document}